\documentclass{IEEEcsmag}
\usepackage[colorlinks,urlcolor=orange,linkcolor=orange,citecolor=orange]{hyperref}
\usepackage{listings}
\expandafter\def\expandafter\UrlBreaks\expandafter{\UrlBreaks\do\/\do\*\do\-\do\~\do\'\do\"\do\-}
\usepackage{booktabs}
\usepackage{numprint}
\npstyleenglish  

\usepackage{multirow}
\usepackage{xspace}
\usepackage{bera}
\usepackage[table]{xcolor}
\usepackage{tikz}
\usepackage{makecell}
\tikzset{
    cross/.pic = {
    \draw[rotate = 45] (-#1,0) -- (#1,0);
    \draw[rotate = 45] (0,-#1) -- (0, #1);
    }
}

\usepackage{silence}
\usepackage{pifont}
\usepackage[frozencache,cachedir=.]{minted}
\AtBeginEnvironment{minted}{\dontdofcolorbox}
\def\dontdofcolorbox{\renewcommand\fcolorbox[4][]{##4}}
\DeclareRobustCommand\redDot{\tikz \fill[blue] (1ex,1ex) circle (0.4ex);}

\newcommand{\cmark}{\ding{51}}%
\newcommand{\xmark}{\ding{55}}%
\newcommand{\revise}[1]{\textcolor{black}{#1}}
\newcommand{\cyclonedxmavenplugin}{\texttt{CycloneDX-Maven-Plugin}\xspace}
\newcommand{\cdxgen}{\texttt{CycloneDX-Generator}\xspace}
\newcommand{\jbom}{\texttt{jbom}\xspace}
\newcommand{\buildinfogo}{\texttt{Build-Info-Go}\xspace}
\newcommand{\depscan}{\texttt{Depscan}\xspace}
\newcommand{\openrewrite}{\texttt{OpenRewrite}\xspace}

\colorlet{punct}{red!60!black}
\definecolor{background}{HTML}{EEEEEE}
\definecolor{delim}{RGB}{20,105,176}
\colorlet{numb}{magenta!60!black}

\jvol{XX}
\jnum{XX}
\paper{8}
\jmonth{June}
\jname{}
\jtitle{}
\pubyear{2023}

\begin{document}

\sptitle{}

\title{Challenges of Producing Software Bill Of Materials for Java}


\author{Musard Balliu,
Benoit Baudry,
Sofia Bobadilla,
Mathias Ekstedt,
Martin Monperrus,
Javier Ron,
Aman Sharma,
Gabriel Skoglund,
César Soto-Valero,
Martin Wittlinger}
\affil{}

\author{\{musard, baudry, sofbob, mekstedt, monperrus, javierro, amansha, gabsko, cesarsv, marwit\}@kth.se}

\begin{abstract}
Software bills of materials (SBOM) promise to become the backbone of software supply chain hardening. We deep-dive into 6 tools and the accuracy of the SBOMs they produce for complex open-source Java projects. Our novel insights reveal some hard challenges regarding the accurate production and usage of software bills of materials.
\end{abstract}

\maketitle

\section{Introduction}

\revise{Modern software applications are virtually never built entirely in-house. As a matter of fact, they reuse many third-party dependencies, which form the core of their software supply chain~\cite{cox2019surviving}. 
The large number of dependencies in an application has turned into a major challenge for both security and reliability~\cite{gkortzis2021software}.
For example, to compromise a high-value application, malicious actors can choose to attack a less well-guarded dependency of the project~\cite{ladisa2022taxonomy}.
Even when there is no malicious intent, bugs can propagate through the software supply chain and cause breakages in applications~\cite{rezk2021ghost}.
Gathering accurate, up-to-date information about all dependencies included in an application is, therefore, of vital importance.}

The Software Bill of Materials (SBOM) has recently emerged as a key concept to enable principled engineering of software supply chains. This takes the well-known concept of `bill of materials' for manufacturing physical goods into the world of software development.
\revise{The purpose of an SBOM is to capture relevant information about the internals of a software artifact. First and foremost, an SBOM is expected to include a complete inventory of all the third-party dependencies of the artifact.}  

Accurate SBOMs are essential for software supply chain management \cite{harutyunyan2020managing}, vulnerability tracking, build tampering detection \cite{nikitin2017chainiac}, and  high software integrity. 
For example, software developers leverage SBOMs to identify vulnerable software components in a timely manner.
This is usually done by matching software component versions against vulnerability databases and reporting a warning whenever a vulnerable component is part of an application.
For example, in 2021, a serious vulnerability present in the popular Java logging component Log4J was discovered.
\revise{This component was extensively used by a large number of open-source and proprietary projects, and consequently, it was a tedious and costly endeavour to identify all impacted projects~\cite{log4j}.
Had all these Java projects published an SBOM, it would have facilitated the precise identification and remediation of vulnerable applications.}

\revise{The software supply chain of modern applications includes hundreds of components, and to have humans producing SBOMs by hand is an unreasonable, time-consuming, and error-prone task.}
Yet, the full automation of SBOM production is a process that poses several challenges \cite{sbom_formats}.
First, the SBOM must elicit all direct dependencies, which are explicitly declared by the application's developers in a build configuration file, as well as the indirect dependencies that come from the transitive closure of dependencies.
\revise{Tracking down every single dependency that is being used is hard when software architectures are formed by deeply nested components, some of which are potentially resolved at runtime.}
Identifying the exact version of a binary dependency in an SBOM is even harder as this requires tracing the binary components back to source code repositories. 
Second, while some package managers are able to list the  dependencies, SBOMs are meant to include extra information about the software supply chain, such as checksums for all dependencies and data about third-party tools used in the build. Finally, the SBOM aims at being both human-readable for auditing and legal cases, as well as machine-readable for automatic verification. 
These challenges open an exciting area for research and innovation, as witnessed by the recent emergence of many SBOM tools supported by diverse open-source communities, start-ups, and big tech companies alike.
From a research perspective, there is a crucial need for laying down systematic foundations of what SBOMs are, and the challenges related to their engineering.

This article presents an in-depth study of SBOM producers in the Java ecosystem. 
Our focus on Java is motivated as follows. 
First, it is one of the top-3 languages in the world by most notable metrics. Second, its mature ecosystem of third-party dependencies, mainly through Maven, is critical in government services, financial services \cite{SotoValeroMB22}, medical infrastructure, and enterprise software systems \cite{MassacciP21}.
Third, SBOM production is intrinsically related to programming language specifics, as it must capture each and every aspect of dependency resolution, compilation, linking, and packaging, all being unique for a given software stack.


For our study, we create a curated selection of 6 mature and actively maintained SBOM producers. We execute each producer on a set of 26 active open-source Java projects.
We observe significant variations in the quality of SBOMs generated by these SBOM producers. \revise{ In particular, they capture a different set of dependencies for the same project}. Based on further manual analysis, we highlight urgent challenges and opportunities to consolidate the state-of-the-art of SBOM production, in order to support thorough security and reliability analyses for software supply chains.

\section{Software Bill of Materials}

\begin{figure*}
    \centering    \centerline{\includegraphics[width=\textwidth]{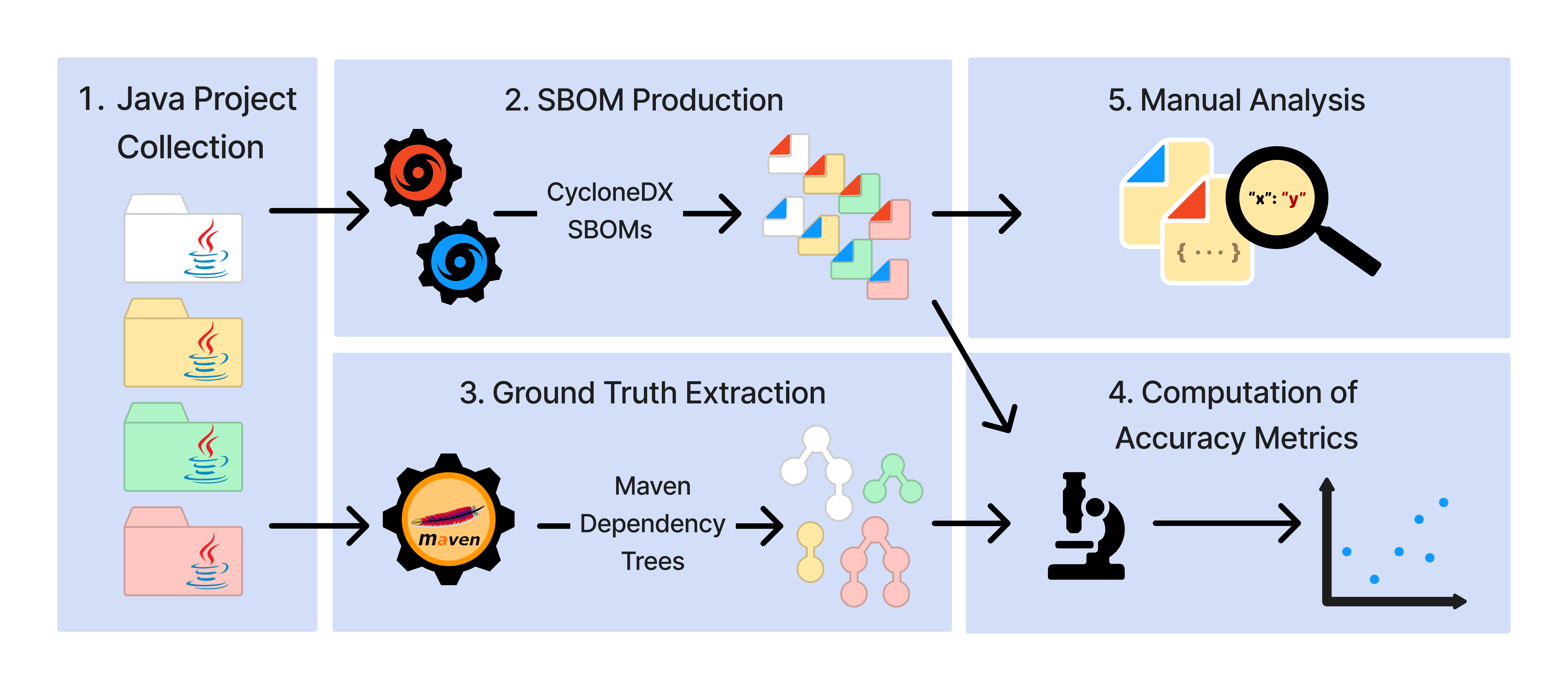}}
    \caption{Overview of 
    the methodology to study CycloneDX SBOM production for Java.}
    \label{fig:THEfigure}
\end{figure*}

\begin{listing}
\begin{minted}[autogobble,
               frame=single,
               framesep=2mm,
               linenos=false,
               xleftmargin=0pt,
               tabsize=1,
               fontsize=\scriptsize,
               breaklines,
               breakanywhere]{json}
{ "bomFormat" : "CycloneDX",
  "specVersion" : "1.4",
  "metadata" : { 
    "timestamp" : "2023-02-20T16:14:42Z",
    "tools" : [
      { "name" : "CycloneDX Maven plugin",
        "version" : "2.7.5" }
    ],
    "component" : {
      "group" : "org.asynchttpclient",
      "name" : "async-http-client-project",
      "version" : "2.12.3",
      "hashes" : [ { "alg" : "SHA-512",
          "content" : "e5435852...7b3e6173"}, ... ]
      "licenses" : [...],
      "externalReferences" : [ {
        "url" : "http://github.com/AsyncHttpClient/async-http-client" }
      ],
      "bom-ref" : "pkg:maven/org.asynchttpclient/async-http-client-project@2.12.3?type=pom"
    }
  },
  "components" : [
    { "group" : "com.sun.activation",
      "name" : "jakarta.activation",
      "version" : "1.2.2",
      "bom-ref" : "pkg:maven/com.sun.activation/jakarta.activation@1.2.2?type=jar"
    } ...
  ],
  "dependencies" : [ {
      "ref" : "pkg:maven/org.asynchttpclient/async-http-client-project@2.12.3?type=pom",
      "dependsOn" : [
        "pkg:maven/com.sun.activation/jakarta.activation@1.2.2?type=jar"
        ....
      ]
    } ... ] }
\end{minted}
\caption{Excerpt of a CycloneDX SBOM for the Java project \texttt{async-http-client}.}
\label{lst:spec}
\end{listing}

In 2021, the United States National Telecommunications and Information Administration~(NTIA) set out to identify a minimal set of requirements for SBOMs~\cite{NTIA21}. These requirements outline which data fields should be present, how SBOMs should support automation, and which practices and processes should be employed when creating, distributing, and using SBOMs. 
The NTIA concluded that three existing formats meet the requirements: CycloneDX, Software Package Data Exchange (SPDX), and Software Identification (SWID).

CycloneDX aims to be a standard for bills of materials for software, hardware, software as a service, and operations. It has a strong security focus, originating from the Open Worldwide Application Security Project (OWASP).
\revise{In this paper, we focus on the CycloneDX standard. This choice is motivated by the rapid development of the standard, as witnessed by the release of many tools for producing CycloneDX SBOMs.}

\autoref{lst:spec} shows an excerpt of a CycloneDX SBOM for the Java component \texttt{async-http-client}. This particular example contains three root elements -- \texttt{metadata}, \texttt{components}, and \texttt{dependencies} -- following the CycloneDX standard. The \texttt{metadata} element records information about the tool which produced the SBOM and the project on which the producer was executed. The \texttt{components} element is a list that includes information about each dependency found in the project.  
Each component's item may also contain hashes to help identify its exact version, which can be used to ensure build integrity. The \texttt{dependencies} element is a list that records the relationship between all of the previously-listed dependencies. In the example, \texttt{jakarta.activation} is  a direct dependency of the analyzed project. 

We remark  that \autoref{lst:spec}  is a simplified SBOM for the sake of clarity. In practice, an SBOM will contain much more data. The full CycloneDX SBOM of \texttt{async-http-client} describes 109 dependencies and provides eight hashes generated through different algorithms for each component (full-fledged real SBOM: \url{https://bit.ly/3lfTdpz}).
Furthermore, the SBOM standard allows recording additional elements such as references to external resources (e.g., the issue tracker), vulnerabilities, and code signatures.

\revise{Given the importance of Java in enterprise and government IT, the production of Java SBOMs is an active area. The critical necessity for grounded and correct Java SBOMs is at the core of this paper's significance.  In what follows, we  purposely produce SBOMs for complex multi-module Java applications, which are archetypal of enterprise Java software systems.}

\section{Methodology to Study SBOM Producers}

\rowcolors{2}{gray!25}{white}
\begin{table*}\centering
\scriptsize
\begin{tabular}{l l || c c c l c}
\toprule
SBOM producer & Version & Checksums & Hierarchy & \revise{Reproducibility} & Production step & Scope \\
\midrule


\href{https://github.com/jfrog/build-info-go}{\buildinfogo} & \revise{1.9.3} & \cmark~(3) & \cmark & \xmark & \revise{Build (Maven compile phase)} & \xmark~(0)  \\
\href{https://github.com/CycloneDX/cdxgen}{\cdxgen} & \revise{8.4.3} & \cmark~(8) & \cmark & \cmark & \revise{Build (Maven package phase)} & \cmark~(1)  \\
\href{https://github.com/CycloneDX/cyclonedx-maven-plugin}{\cyclonedxmavenplugin} & \revise{2.7.8} & \cmark~(8) & \cmark & \cmark & \revise{Build (Maven package phase)} & \cmark~(1)  \\
\href{https://github.com/AppThreat/dep-scan}{\depscan} & \revise{4.1.2} & \cmark~(8) & \cmark & \cmark & \revise{Source (Static source code)} & \cmark~(1)  \\
\href{https://github.com/Contrast-Security-OSS/jbom}{\jbom} & 1.2.1 & \cmark~(2) & \xmark & \xmark & \revise{Analyzed (Post maven package phase)} & \cmark~(1)  \\
\href{https://docs.openrewrite.org/reference/rewrite-maven-plugin}{\openrewrite} & \revise{4.45.0} & \xmark\,~(0) & \cmark & \cmark & \revise{Build (Maven package phase)} & \cmark~(2)  \\
\bottomrule
\hfill
\end{tabular}
\caption{Curated set of SBOM producers subject to our study, supporting Java and the CycloneDX standard.}

\label{tab:sbom-tools}
\end{table*}

The core of our study consists in curating and executing state-of-the-art SBOM production tools on a set of mature Java projects. Then, we perform a comparative analysis of the SBOMs following the methodology illustrated in \autoref{fig:THEfigure}.

\subsection{SBOM producers}
To curate the list of SBOM producers, we start by identifying producers targeting CycloneDX SBOMs for Java projects. We scan through all the candidates from the official \href{https://cyclonedx.org/tool-center/}{CycloneDX tool center} and query GitHub with the keyword 'SBOM' for projects with at least 100 stars. This process yields 24 producers.

\revise{We further select the producers that meet the following criteria. Each selected producer should: 1) produce an SBOM containing the dependencies of the project; 2) be able to analyze Java projects that build with Maven; 3) be open-source; 4) be run as a command-line tool and not only as an online tool. The last two criteria are essential for automating our experiments and for reproducible science.}

Ultimately, this process results in a curated set of 6 SBOM producers,   \buildinfogo, \cdxgen, \cyclonedxmavenplugin,  \depscan, \jbom, and \openrewrite, as shown in \autoref{tab:sbom-tools}.
\revise{We use all of these producers' most recent stable releases as of May 5, 2023.}

\subsection{SBOM Conceptual Framework}
\revise{After a deep analysis of the considered SBOM producers, we postulate the following framework of SBOMs that we will apply to our experimental results.}

\noindent\revise{\emph{Build integrity:} SBOMs can contain checksums of software components for verifying build integrity, but the format of checksums is open.}

\noindent\revise{\emph{Dependency hierarchy:} SBOMs can contain either a flat list of dependencies or structured trees of dependencies, which impacts subsequent consumption.}

\noindent\revise{\emph{Production step:} SBOMs can be computed at different stages of the build and deploy lifecycle, and this can change the resulting SBOMs significantly.}

\noindent\revise{\emph{Dependency resolution:} SBOMs must faithfully capture the dependency resolution as it happens in build tools, which is often not documented.}

\subsection{Projects under study}

\revise{To compare the quality of SBOMs generated by different producers, we run them on a dataset of Java projects. This dataset is meant to include mature, active Java projects that rely on a significant number of dependencies.} A recent work on dependency management in Java has curated a list of projects that meet these criteria  \cite{depclean}. Since our work also involves dependency analysis, we decide to reuse their dataset of projects. The dataset includes 31 Maven projects with stable releases and frequent activity, indicating the project's maturity. 
We exclude \texttt{teavm} 
and \texttt{moshi} 
as these projects have migrated from Maven to using Gradle as the build system, as well as \texttt{auto} and \texttt{subzero} since they are not valid Maven projects due to the lack of a \texttt{pom.xml} file in their root directory.
We merge \texttt{jenkins-core} and \texttt{jenkins-cli} as a single project \texttt{jenkins} as we execute SBOM producers at the root directory of the project rather than submodules to avoid dependency resolution errors. 
This process gives us a set of 26 popular, actively maintained open-source Maven projects for our analysis.

Table~\ref{tab:descriptive-stats} details the set of analyzed Java projects.
Each project is identified by the name and commit at which we analyze the project.
The projects include between 733 and 1.5 million lines of application code and are composed of up to 211 Maven modules. 
They have between 2 and 191 direct dependencies, and between 1 and 582 indirect dependencies.  

\begin{table}
\centering
\scriptsize
\setlength{\tabcolsep}{4.5pt}
\rowcolors{2}{gray!25}{white}
\begin{tabular}{l r r r r r}
\toprule
Project Name &
kLOC &
{\makecell{Maven \\ Modules}} &
DD & 
ID &
Total \\

\midrule
\href{https://github.com/apache/tika/tree/41319f3c294b13de5342a80570b4540f7dd04a3e}{tika} & \numprint{163} & 108 & 186 & 563 & 749 \\ 
\href{https://github.com/Alluxio/alluxio/tree/d5919d8d80ae7bfdd914ade30620d5ca14f3b67e}{alluxio} & \numprint{295} & 66 & 143 & 582 & 725 \\ 
\href{https://github.com/jooby-project/jooby/tree/f71b551213ac03523e44a7fbb8c972b752ffc707}{jooby} & \numprint{65} & 54 & 129 & 368 & 497 \\ 
\href{https://github.com/neo4j/neo4j/tree/c082e80b792d46ad1b342fbf7f1facb2028344c6}{neo4j} & \numprint{686} & 124 & 191 & 273 & 464 \\ 
\href{https://github.com/apache/flink/tree/c41c8e5cfab683da8135d6c822693ef851d6e2b7}{flink} & \numprint{1528} & 211 & 121 & 270 & 391 \\ 
\href{https://github.com/eclipse/steady/tree/3d261afe9513f7c708324aa0183423ab2e9e4692}{steady} & \numprint{99} & 20 & 78 & 267 & 345 \\ 
\href{https://github.com/Erudika/para/tree/41d900574e2e159b05fbd23aaab1f6e554ab8fc3}{para} & \numprint{29} & 6 & 82 & 224 & 306 \\ 
\href{https://github.com/jenkinsci/jenkins/tree/ce7e5d70373a36c8d26d4117384a9c5cb57ff1c1}{jenkins} & \numprint{181} & 10 & 99 & 200 & 299 \\ 
\href{https://github.com/apache/accumulo/tree/706612f859d6e68891d487d624eda9ecf3fea7f9}{accumulo} & \numprint{399} & 18 & 121 & 158 & 279 \\ 
\href{https://github.com/vmi/selenese-runner-java/tree/3e84e8e4e7e06aa1bdacaa8266db00f62ebef559}{selenese-runner-java} & \numprint{21} & 1 & 22 & 114 & 136 \\ 
\href{https://github.com/undertow-io/undertow/tree/f52b70c1520277a1552f0f453c2a908897a8a5dc}{undertow} & \numprint{150} & 10 & 28 & 107 & 135 \\ 
\href{https://github.com/jknack/handlebars.java/tree/2afc50fd5dcd32af28f8305b59689b3fec4a3b07}{handlebars.java} & \numprint{22} & 11 & 36 & 84 & 120 \\ 
\href{https://github.com/google/error-prone/tree/27de40ba6008f967c01a55ec83c9127419bfe433}{error-prone} & \numprint{225} & 10 & 61 & 53 & 114 \\ 
\href{https://github.com/AsyncHttpClient/async-http-client/tree/7a370af58dc8895a27a14d0a81af2a3b91930651}{async-http-client} & \numprint{29} & 14 & 40 & 69 & 109 \\ 
\href{https://github.com/rnewson/couchdb-lucene/tree/855473709bd4e3d92d3f62ece86ab739d0f0de13}{couchdb-lucene} & \numprint{3.9} & 1 & 25 & 51 & 76 \\ 
\href{https://github.com/mybatis/mybatis-3/tree/c195f12808a88a1ee245dc86d9c1621042655970}{mybatis-3} & \numprint{62} & 1 & 27 & 37 & 64 \\ 
\href{https://github.com/orphan-oss/launch4j-maven-plugin/tree/3f9818ee34b36cdcea58e2d6e6542f140b394faf}{launch4j-maven-plugin} & \numprint{1.5} & 1 & 12 & 50 & 62 \\
\href{https://github.com/checkstyle/checkstyle/tree/233c91be45abc1ddf67c1df7bc8f9f8ab64caa1c}{checkstyle} & \numprint{304} & 1 & 22 & 35 & 57 \\ 
\href{https://github.com/orika-mapper/orika/tree/eef82092c8a9dfda04192a5378fa0e49d70ade3a}{orika} & \numprint{43} & 5 & 25 & 30 & 55 \\ 
\href{https://github.com/apache/commons-configuration/tree/59e5152722198526c6ffe5361de7d1a6a87275c7}{commons-configuration} & \numprint{51} & 1 & 33 & 21 & 54 \\ 
\href{https://github.com/INRIA/spoon/tree/ee73f4376aa929d8dce950202fabb8992a77c9fb}{spoon} & \numprint{155} & 1 & 22 & 32 & 54 \\
\href{https://github.com/sarxos/webcam-capture/tree/e19125c2c728a856231a3b507372e94e02fdfd35}{webcam-capture} & \numprint{19} & 2 & 16 & 35 & 51 \\ 
\href{https://github.com/javaparser/javaparser/tree/1ae25f3f77f5d680c135d0742257ccd62916f17d}{javaparser} & \numprint{181} & 11 & 18 & 33 & 51 \\ 
\href{https://github.com/stanfordnlp/CoreNLP/tree/f7782ff5f235584b0fc559f266961b5ab013556a}{CoreNLP} & \numprint{615} & 3 & 23 & 18 & 41 \\ 
\href{https://github.com/radsz/jacop/tree/1a395e6add22caf79590fe9d1b2223bfb6ed0cd0}{jacop} & \numprint{89} & 1 & 6 & 5 & 11 \\ 
\href{https://github.com/giltene/jHiccup/tree/a440bdaed143e1445cbeab7c5bffd30989a435d0}{jHiccup} & \numprint{0.7} & 1 & 2 & 1 & 3 \\   
\bottomrule
\hfill
\end{tabular}
\caption{Descriptive statistics of the analyzed Java projects. Number of thousands of lines of application code (kLOC), number of Maven modules, number of unique direct dependencies (DD), number of unique indirect dependencies (ID), and total number of unique dependencies (Total). Rows are ordered with respect to the total number of dependencies in the projects.}
\label{tab:descriptive-stats}
\end{table}


\subsection{Protocol to compare SBOM producers}
\autoref{fig:THEfigure} illustrates the five main steps of the protocol for our experiment.
\revise{ Step 2 in ~\autoref{fig:THEfigure} is ``SBOM Production'' where we run each SBOM producer on each project. }
To support the reproducibility of our experiment, we save the specific git hash of each project and run the SBOM producers in a docker container.
This SBOM generation procedure is fully automated, and it ensures that there are no interactions among the producers as the SBOM production for each project is isolated and starts in the same state.
The repository with our study subjects and the experimental pipeline is publicly available\footnote{\url{https://github.com/chains-project/SBOM-2023}}.

An SBOM captures a rich set of information about the software supply chain, including the network of direct and indirect dependencies. As part of our study, we  assess the accuracy of the dependencies in the SBOM with respect to a ground truth. \revise{ Step 3 in ~\autoref{fig:THEfigure} represents the process of extracting the ground truth. We use the complete list of dependencies returned by the command \texttt{tree} of the \texttt{maven-dependency-plugin@3.4.0}.}
This plugin is an integral part of the Maven build system, and it is the most common plugin used to perform this single task in the supply chain: resolve dependencies. It provides a deterministic dependency tree for a specific version of a Maven project. Moreover, it has been in production since 2007 and is being continuously maintained, with the latest release as recent as 2023. It is very mature and stable, and consequently is the best ground truth for our study.

In Maven, a dependency is identified by a name and a version number. The name is a combination of its \texttt{groupId} and \texttt{artifactId}, separated by a colon, for example, \texttt{com.google.guava:guava}. We consider two dependencies identical if their name and version match precisely. \revise{As shown in step 4 in~\autoref{fig:THEfigure}, we compare the accuracy of SBOMs by computing the precision and recall of dependency lists computed by each producer}. 
The precision is the share of dependencies in the SBOM that are correct with respect to the ground truth.
The recall is the share of correct dependencies that are in the SBOM.

\revise{Note that the ground truth considers all dependencies required for producing a software artifact, including test dependencies. While these test dependencies are not included in the deployed software, they are relevant in the context of supply chain attacks. A malicious test dependency has the potential to interact with the build system and introduce malicious code at build time~\cite{ladisa2022taxonomy}. In order to trace vulnerable or malicious test dependencies, it is important that these are included in the SBOM.} 

The last step of our methodology\revise{, step 5 in~\autoref{fig:THEfigure},} consists of manually analyzing a sample of SBOMs in order to get a concrete grasp at the content of the SBOMs produced. This provides us with detailed insights about the challenges that SBOM producers face in order to correctly retrieve all the dependencies in an application's software supply chain.

\section{Experimental Results}

We follow our protocol and run 6 SBOM production tools on 26 Java projects.  The results provide key insights about the tools' behavior as well as the quality of the produced SBOMs.

\subsection{Producer Insights}

\autoref{tab:sbom-tools} summarizes the essential features of SBOM producers that we have identified, and to what extent these features are present in the tools.

\paragraph{Checksum Diversity}
\autoref{tab:sbom-tools} summarizes the number of different checksum algorithms that each SBOM producer uses. 
The production of different checksums is useful because it maximizes the likelihood of integrating the SBOMs with third-party tools that expect a specific checksum.
Three producers compute eight types of checksums for each dependency jar:
\cdxgen, \cyclonedxmavenplugin and \depscan provide md5, sha1, sha256, sha512, sha384, sha3-384, sha3-256, and sha3-512 for each dependency in the SBOM. 
One producer, \openrewrite, does not provide any checksums, which is considered a serious limitation. 
Our observations help practitioners to select SBOM producers accordingly.

\paragraph{Dependency Hierarchy}

An essential feature of SBOM producers is eliciting all the dependencies in the software supply chain of an application.
Beyond a flat list, some analyses, such as vulnerability analysis, debloating, and installation via package managers, require the complete tree of relationships between the different components in the chain.
The CycloneDX specification provides the attribute \texttt{dependencies} to serve this purpose.
We note that five of six producers report the hierarchy among dependencies.
However, \jbom cannot link the dependencies together since it acts after the build step where some dependencies cannot be resolved.
For example, for \texttt{mybatis-3},  \texttt{com.fasterxml.jackson.core:jackson-core} version \texttt{2.13.2} is an indirect dependency at the fourth level.
The producers \buildinfogo, \cdxgen, \cyclonedxmavenplugin, and \depscan report this information correctly.

\paragraph{\revise{ Reproducibility}}
SBOMs are meant to be reference documents, and potentially may become legally binding. To that extent, one must produce them reliably.
\revise{In that respect, we claim that SBOM production should be reproducible. We say an SBOM tool is reproducible if it generates strictly identical files contentwise over multiple runs. We exclude metadata such as timestamp.
We generate SBOMs twice for each producer and find that \buildinfogo and \jbom are not reproducible: they do not preserve the order of SBOM elements. Moreover, \jbom also produces different hashes of the components. While this is a fixable engineering issue, it highlights the necessity to consolidate the maturity of SBOM tooling before it can be relied upon in court.}

\paragraph{Production Step}

\revise{There are six steps at which an SBOM could be produced - Design, Source, Build, Analyzed, Deployed, and Runtime\footnote{\url{https://www.cisa.gov/sites/default/files/2023-04/sbom-types-document-508c.pdf}}. The considered SBOM producers do not produce SBOM at the same step. We report the step at which SBOM is produced per the documentation provided by the developers. \buildinfogo, \cdxgen, \cyclonedxmavenplugin, and \openrewrite produce an SBOM at the Build step. The build step can further be broken down into more steps, as Maven splits a build into multiple phases\footnote{\url{https://maven.apache.org/guides/introduction/introduction-to-the-lifecycle.html}}. \buildinfogo produces an SBOM when the Maven build system is compiling. Meanwhile, the other three producers perform SBOM production when the artifact, JAR, for example, is being generated. This phase is also called \texttt{package} in  Maven. \depscan produces an SBOM from the source files. Finally, \jbom produces an SBOM by analyzing the final jar file, corresponding to CISA's ``Analyzed'' step.}	

These different steps are significant regarding the production of SBOMs, since the information available about the software supply chain varies at these different stages.
Indeed, software projects go through a Build/\linebreak[4]CI/CD life cycle  and, at every point, the information available is different~\cite{xia2023}.
For example, before the build phase, an SBOM producer cannot know what will be finally included in the binary.
Similarly, after the build,  information about some dependencies may be lost, because the build system has removed redundant or unnecessary dependencies.

The CycloneDX standard does not address this aspect, and the producers do not clearly document or motivate the phase they consider. SBOM producers should state the production step at which they collect information about the software supply chain, to help SBOM consumers decide which SBOM is most appropriate for their needs.

\paragraph{Scopes}\label{para:sbom_scope_problem}

The CycloneDX JSON specification supports an optional \texttt{scope}  attribute for each component. This attribute can take the values \texttt{required}, \texttt{optional} or \texttt{excluded}, based on the dependency's behavior at runtime. According to the specification\footnote{\url{https://github.com/CycloneDX/specification/blob/1.4/schema/bom-1.4.xsd\#L514}}, the \texttt{required} scope denotes that the component is required at runtime; the \texttt{optional} scope denotes components that \textit{``[\dots] are not capable of being called due to them not be installed or otherwise accessible by any means''}. Finally, \texttt{excluded} components \textit{``[\dots] provide the ability to document component usage for test and other non-runtime purposes.'}

We observe significant differences among SBOM producers regarding the identification of scopes. 
\revise{As an example, \texttt{org.slf4j:slf4j-api@2.0.1} is a dependency of \texttt{mybatis-3}. \cdxgen,  \cyclonedxmavenplugin, and \depscan report its scope as \texttt{optional}, \openrewrite reports the scope as \texttt{required}, while the other producers report no scope at all}. 

We note that each SBOM producer only uses a subset of the allowed scope values. \cdxgen, \cyclonedxmavenplugin, and \depscan either label components as \texttt{optional} or provide no scope value. \jbom labels all components as \texttt{required}. \openrewrite marks components as either \texttt{optional} or \texttt{required}, and \buildinfogo does not report \texttt{scope} for any component. It is not clear from the documentation of the producers how these values are computed.
Due to the lack of clarity in the standard and the absence of ground truth, it is impossible to determine which one is correct.

Providing clear information as to how and when in the software lifecycle a component is used -- the scope as we understand the standard -- is an important feature of an SBOM. However, our results show that no SBOM consumer can rely on the \texttt{scope} values produced by current SBOM producers.

\subsection{Dependency Identification Accuracy}

\begin{figure}
  \begin{center}
    \includegraphics[scale=0.5]{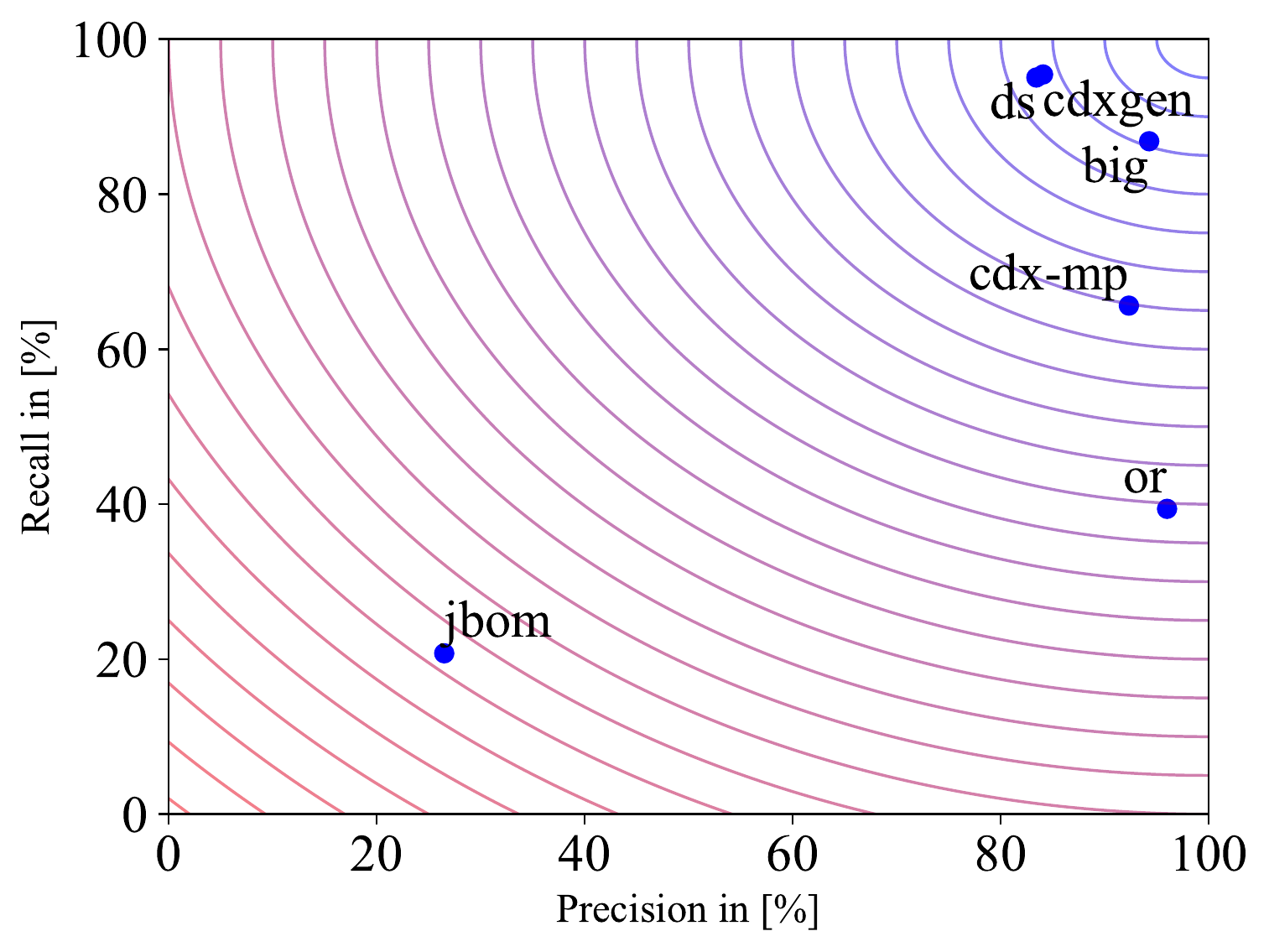}
    \caption{\revise{Mean precision and mean recall of each SBOM producer, excluding producer failures.}
    The abbreviations stand for the following, \texttt{big} for \buildinfogo, \texttt{cdx} for \cdxgen, \texttt{or} for \openrewrite and \texttt{cdx-mp} for \cyclonedxmavenplugin}
    \label{fig:sbomProducerResults}
  \end{center}
\end{figure}

\revise{\autoref{fig:sbomProducerResults} shows the accuracy of the considered SBOM producers per our ground truth. The X-axis is the precision, and the Y-axis is the recall for each producer. A point \redDot \ represents the accuracy of dependencies captured in the SBOM and the isolines represent the standard F1-score combining precision and recall}.  We report the average precision and recall of an SBOM producer, over all projects. 
For our experiment, we performed 156 executions of SBOM producers, which produced  \revise{119} SBOMs and \revise{37} failures. For the latter, SBOMs were either empty or contained no dependency because the build failed or the producer failed. We exclude these data points from our study.

\revise{At the bottom-left of  \autoref{fig:sbomProducerResults}, \jbom has the lowest precision and recall. Next, \openrewrite has the highest  precision of 96\%, but with a low recall. Higher up in the figure, we find \cyclonedxmavenplugin with 92\% precision and 66\% recall; \cdxgen and \depscan perform very similarly. Finally, \buildinfogo is at the top right corner with the best score of detected dependencies according to the dataset and ground truth.}

\revise{We highlight five main reasons why producers fall short on creating a fully-accurate SBOM with respect to  the ground truth:  exclusion of test dependencies from the SBOM; failure to resolve maven properties\footnote{\url{https://maven.apache.org/pom.html\#Properties}}; failure to correctly resolve the \texttt{version} of a dependency;  advanced dependency resolution techniques;  the project itself is counted as a dependency. We elaborate on each of these points below.}

\revise{SBOM producers like \openrewrite and \cyclonedxmavenplugin do not include test dependencies by default in the SBOM they produce. This explains the low recall of 39\% and 66\%, respectively. Although \buildinfogo has the highest F1-score, we observe that it  misses test dependencies for some projects, while achieving 100\% recall on some other projects, for example \texttt{jenkins}, which clearly contains test dependencies.}

\revise{When a producer does not correctly resolve Maven properties,  the  SBOM cannot be compared to the ground truth. For example, \jbom reports version \texttt{\$\{guava.version\}} for \texttt{com.google.guava:guava}, instead of  \texttt{31.0.1-jre}, for \texttt{alluxio}. This eventually yields a list of dependencies that are not comparable with the list of dependencies in the ground truth.}

\revise{To verify that a dependency is correctly reported, the \texttt{groupId}, \texttt{artifactId}, and \texttt{version} must match. However, \jbom incorrectly  retrieves the \texttt{version} for some dependencies. For example, it reports version \texttt{0.4}  of \texttt{com.pholser:junit-quickcheck-core} for \texttt{CoreNLP}, which does not exist in the ground truth. \openrewrite faces similar challenges, reporting version \texttt{4.1.78.Final} for \texttt{io.netty:netty-handler} in \texttt{selenese-runner-java}, while the correct version is \texttt{4.1.79.Final}.
A major difficulty for retrieving the version number occurs when different versions of the same library appear as indirect dependencies at different locations in the dependency tree. The correct version identification must faithfully capture the actual resolution embedded in the build system.}

\revise{Moreover, the resolution of dependencies is affected by the different ways that SBOM producers use to retrieve dependencies.  \depscan and \cdxgen perform equally on most projects. For example, they both have the same results on \texttt{selenese-runner-java}. However, \depscan correctly reports a dependency \texttt{ch.uzh.ifi.seal:changedistiller} version \texttt{0.0.4} for \texttt{steady} while \cdxgen misses it. In this case, \depscan reports a dependency that is stored as local jar in the project. This illustrates that they both have different methods for resolving dependencies. On the other hand, a closer look  at the architecture of \buildinfogo shows that it relies on the Maven APIs to invoke the build and retrieve deeper information, thus producing results that are closer to the  ground truth.
This suggests that SBOM producers  benefit from being tailored to a language and a build system  in order to plug deeply into the build process to obtain correct information.}

\revise{Finally, we have observed that some SBOMs include the source project in the dependency list. This is yet another reason why \buildinfogo falls short of perfect aligment with the ground truth. It reports all dependencies for \texttt{selenese-runner-java}, \texttt{accumulo}, \texttt{jenkins}, \texttt{checkstyle}, \texttt{error-prone}, \texttt{jooby}, \texttt{launch4j-maven-plugin}, \texttt{orika}, and \texttt{mybatis-3}, but in each of these cases, it incorrectly considers the root module as a dependency.}


Overall, \autoref{fig:sbomProducerResults} shows significant differences among the  accuracy of the SBOMs produced by 6 state of the art producers. These results  reveal discrepancies in the list of dependencies  in the SBOMs, with different dependency versions and missed dependencies.
To better illustrate   the different accuracy levels, we manually analyze a sample of the SBOMs. \revise{To sample the files we use the following criteria: Ww select SBOMs produced by \buildinfogo and \jbom as these  producers are at both ends of the accuracy range; we analyze SBOMs for project \texttt{spoon} as three of the authors are maintainers and hence have a deep understanding of this project.
After applying the previous filters we sample 4 SBOM files: 2 SBOMs with the highest and lowest precision on dependencies, produced by \buildinfogo and \jbom; 
2 SBOMs with the highest and lowest precision on direct dependencies produced.}
This analysis was conducted by two of the authors, both experts in Java programming. In case of discrepancies, they met and discussed to resolve them and reach a conclusion.

The ground truth indicates that the single module of \texttt{spoon} has 22 direct dependencies and 32 indirect ones (see \autoref{tab:descriptive-stats}).
The SBOM produced by \buildinfogo correctly contains \revise{23 dependencies, and the only incorrect one is the \texttt{fr.inria.gforge.spoon:spoon-core} itself}. The precision is consequently high, but some dependencies are clearly missing. \buildinfogo excludes test dependencies for \texttt{spoon}.
\revise{On the other hand, the SBOM produced by \jbom reports 125 dependencies, but only 29 of them are correct}.
\revise{The other 96  dependencies are the result of failure of \jbom to resolve Maven properties, versions or metadata \texttt{groupId}.}

The next two case studies come from \buildinfogo. First, we inspect the SBOM produced for \texttt{selenese-runner-java}, and we find that \buildinfogo fetches all \revise{136} dependencies. It also includes the complete dependency tree hierarchy information. Such precise information is important and makes the SBOM consumable. However, we notice that even a solid producer such as \buildinfogo does not always achieve high precision. \revise{For example, the SBOM of \texttt{javaparser} includes 14 correct dependencies out of 51. The majority of the dependencies the producer misses are test scoped. We observe an inconsistent behavior in \buildinfogo as it sporadically includes test dependencies.}

\revise{The SBOM produced by \jbom for \texttt{async-http-client} contains only 2 correct dependencies out 109.}
On a deeper inspection, we observe that most dependencies in the SBOM are identified with wrong versions, resulting in poor precision. We analyze the SBOM of \texttt{mybatis-3} produced by \jbom.
\revise{This SBOM includes all the direct dependencies, precisely with correct version numbers as they were specified.
However, all indirect dependencies are missed.}

\begin{figure}
    \includegraphics[scale=0.20]{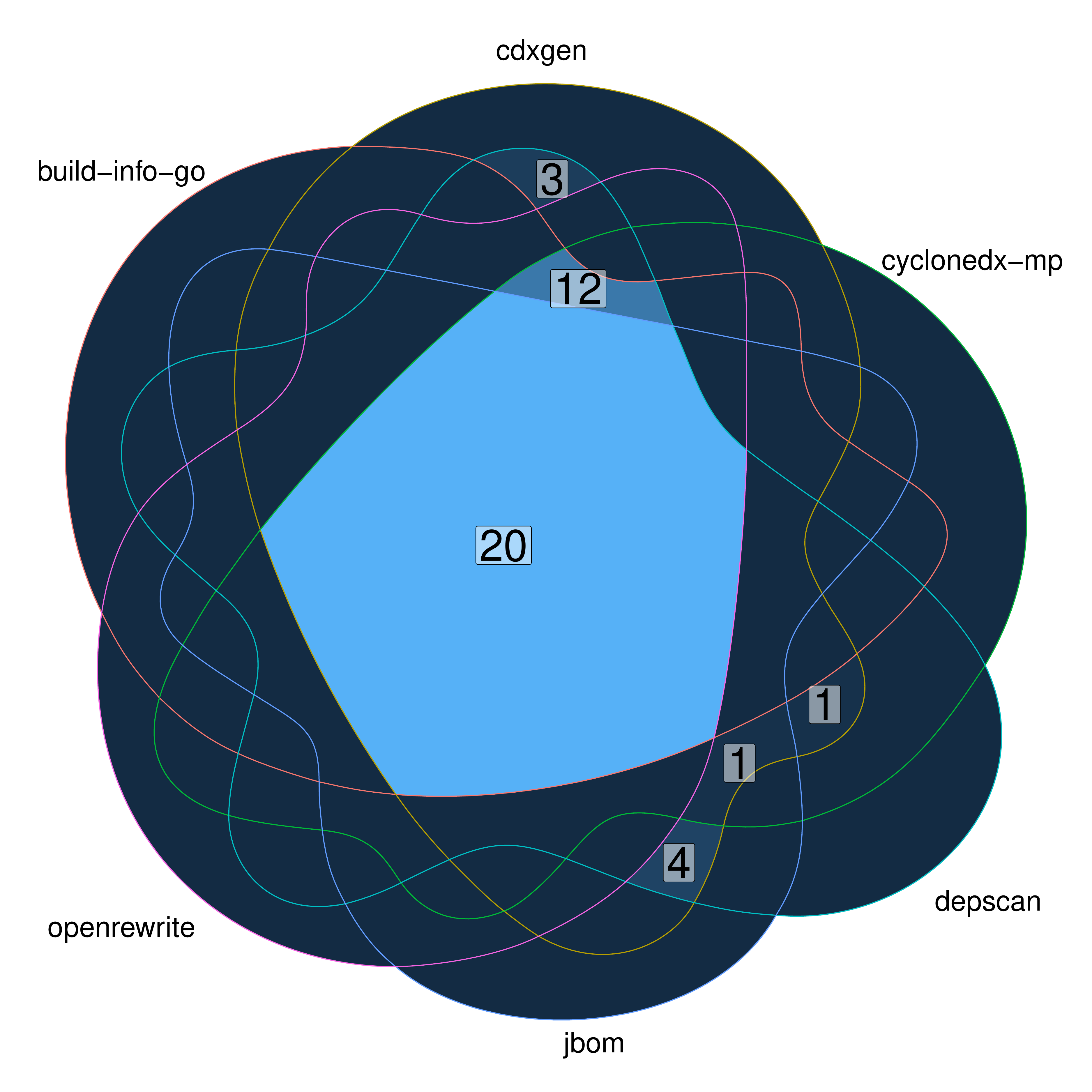}
    \caption{Venn Diagram of different SBOM producer results. Only the true positives (correctly identified dependencies)  are compared. Intersection areas mean that multiple SBOM producers have overlapping correct dependencies. In this project, all producers correctly identify a majority of 20 dependencies.}
    \label{fig:venn}
\end{figure}

\revise{
\emph{Overlap analysis. } 
\autoref{fig:venn} is a Venn diagram that captures the overlap between the SBOMs of  \texttt{CoreNLP} generated by  the six SBOM producers.
For each SBOM producer, we use the set of true positives dependencies.
Intersection areas mean producers have the same correct dependency in their SBOM.
Every SBOM producer has a different color for their outline.
For example, we use yellow for \cdxgen. 
The labels indicate the number of dependencies in the intersection area and areas without a label are empty (meaning no dependency in common).
We have six different intersection areas.
The largest one is in the middle and shows that 20  dependencies are correctly identified by every producer.
The second-largest area indicates that 12 dependencies are correctly captured by every producer except \jbom. For example,
\jbom misses \texttt{javax.xml.bind:jaxb-api} because it either resolves an incorrect version, or it resolves  some dependencies as \texttt{null}.
Two areas have only one dependency in the intersection.
One intersection area contains the producers \cdxgen, \cyclonedxmavenplugin, \jbom, and \depscan, which correctly capture \texttt{junit:junit:4.13.1}, while 
\openrewrite and \buildinfogo miss it.
\openrewrite entirely skips test dependencies by design, and \buildinfogo misses it.
The other area is the intersection of \cyclonedxmavenplugin,\cdxgen, and \depscan that correctly detect \texttt{org.hamcrest:hamcrest-core:1.3} in the SBOM.
\jbom misses this dependency because it is included as a jar with a relative path in the repository.
It only identifies correctly the groupId and artifactId, while the version is set to null.
}

\subsection{Experimental Limitation}

\revise{It may be argued that SBOM producers should simply reuse the ground truth we consider as the basis for SBOM production, that is the code of Maven in our experiment.
However, SBOMs can be extracted at multiple steps, per our discussion on production steps above. All these extraction steps are valid and potentially useful depending on the goal and the SBOM consumption. Our ground truth  only captures one single production step. To that extent, some inaccuracies we have reported may be due to the mismatch between the ground truth and the targeted production steps of some SBOM producers.}

\subsection{Take-aways}

\revise{In theory, extracting SBOMs is easy. Our results show that in practice, this is not the case. In this section, we discuss the benefits of our work for two target audiences, Java developers and standardization committees, and reason about the difficulty of confronting theory and practice.}

\emph{Java developers:}
Our in-depth study shows that \buildinfogo is the best SBOM producer for Java developers.
\revise{The reasons are that:
1) it produces different checksums ; 2) it supports dependency hierarchies; and  
3) it achieves the highest precision and recall thanks to a tailored integration in Maven. }
\revise{Yet, \buildinfogo has room for improvement. First, the precision and recall of 94\% and 87\% respectively can be increased, with several important fixes. Second, assuming that the standard clarifies the matter, it could also provide the scope of the dependencies.}

\emph{Standardization committees:} 
Our study identifies two shortcomings in the CycloneDX standard.
The specification needs to require producers to specify the exact step at which the SBOM is produced, and it must precisely define the notion of scope, which would help both SBOM producers and consumers. 
We believe that the latter is more important as the current state is ambiguous for developers, and ambiguity upstream typically means incorrectness downstream.

\emph{Difficulty:} 
\revise{Our study reveals difficulties of different nature in producing  complete and useful SBOMs.
The challenges of checksums, tree hierarchy, and determinism can all be fixed with additional engineering effort. However, clarifying the meaning of production steps and scopes is fundamentally hard, because it requires the appropriate abstraction over multiple build pipelines in different software stacks, and this abstraction would require consensus in the SBOM community.}

\section{Open Challenges}

Our experiments revealed a number of challenges for  the accurate production  and the effective consumption of Software Bill of Materials.

\paragraph{SBOM and Tooling Dependencies}

In our analysis, we observe that the bulk of SBOMs consists of 
collecting accurate dependency trees for an application project. Yet, the software supply chain of an application is made of many more components. For example, the version control system, the testing and build tools  and the infrastructure to deploy or distribute the application are key components of the supply chain. 
In the recent years, we witnessed   attacks such as the Solarwinds incident, which successfully compromised a system through these components~\cite{ladisa2022taxonomy}. 
The CycloneDX standard attempts to document such information by providing the attribute \texttt{externalReferences}. However, there is currently scarce support to generate these attributes and our study shows that the SBOM producers implement this partially and with inconsistencies. The comprehensive collection and documentation of all tools involved in the supply chain is a pressing challenge to produce SBOMs that are amenable to thorough hardening procedures.

\paragraph{SBOMs for Threat Analysis}

In the longer term, the value of SBOMs will increase with enabling automatic security analyses. For example, one key challenge is to let SBOM producers qualify the trust that one can have in the dependencies. This type of assessment of the supply chain relates to threat modeling and analysis, which is already considered good practice for DevOps organizations \cite{rafi2021readiness}.  In order to guide which properties an SBOM should include support reasoning about trust and threats, the attack taxonomy of Ladisa et al. \cite{ladisa2022taxonomy} constitutes an excellent starting point. Furthermore, the work of Zahan et al. \cite{zahan2022} proposes concrete metrics as warning signs of supply chain vulnerabilities that could be mapped to the taxonomy, such as \textit{Too many maintainers} which can match the \textit{Take-over Legitimate Account} as well as the use of \textit{Installation scripts} which relates to the \textit{Running a malicious build job} technique. 

\paragraph{SBOMs at Runtime}

The next challenge will be to bring SBOMs online, as a foundation to enforce security requirements at runtime. For a given SBOM pertaining to a software application, one can develop lightweight dynamic analysis to enforce mandatory access control policies.
This can be achieved by monitoring the usage of dependencies at runtime and ensuring that only the dependencies within the SBOM are used by the application, thus preventing the entire class of vulnerabilities that rely on the dynamic inclusion of malicious code and packages. A major challenge for such an approach is that it would require accurate static information about dependencies, which is a challenging endeavor, as we have shown in this article.

\paragraph{SBOMs in Other Software Stacks}
\revise{The production of SBOMs for other software stacks is likely to face similar challenges as those seen for Java. We note that some programming ecosystems already partially address certain challenges.
For instance, ecosystems such as npm, Go and Rust record checksums for all publicly available dependencies in auto-generated lock files.
In theory, the data provided by these instruments can already be aggregated and used to produce meaningful SBOMs.
In the specific case of Go, the lock file information can be validated against an immutable, verifiable database, providing integrity guarantees which can be leveraged in SBOMs.
Nonetheless, a definitive solution is yet to be established and widely used in either of these  software stacks.}

\section{Conclusion}

We performed a deep-dive into the meaning of Software Bill of Materials and its realization in the Java ecosystem, one of the most commonly used enterprise programming languages.
Our research findings indicate strong interest and vibrant activity in this essential area for software supply chain security and reliability. Yet, we also revealed that SBOMs today rely on a technical foundation that is unstable.
Our empirical insights shed light on important weaknesses that require attention, starting with incorrect or incomplete dependency lists recovered in SBOMs. 
These findings call for further work in clarifying the SBOM standards, as well for more work on improving the quality of SBOM producers' output. Studying SBOM quality for other languages (e.g. Rust) and for other SBOM formats (e.g. SPDX) would be very valuable for the community. 
Both academia and industry agree that SBOMs promise great benefits, now the time is ripe to all work together to unleash their full potential.
\IEEEtriggeratref{8}
\newpage
\bibliographystyle{ieeetr}
\bibliography{main}

\begin{thebibliography}{10}

\bibitem{cox2019surviving}
R.~Cox, ``{Surviving Software Dependencies},'' {\em Communications of the ACM},
  vol.~62, no.~9, pp.~36--43, 2019.

\bibitem{gkortzis2021software}
A.~Gkortzis, D.~Feitosa, and D.~Spinellis, ``{Software reuse cuts both ways: An
  empirical analysis of its relationship with security vulnerabilities},'' {\em
  Journal of Systems and Software}, vol.~172, 2021.

\bibitem{ladisa2022taxonomy}
P.~Ladisa, H.~Plate, M.~Martinez, and O.~Barais, ``{SoK: Taxonomy of Attacks on
  Open-Source Software Supply Chains},'' in {\em Proceedings of the IEEE
  Symposium on Security and Privacy (SP)}, may 2023.

\bibitem{rezk2021ghost}
C.~Rezk, Y.~Kamei, and S.~Mcintosh, ``{The ghost commit problem when
  identifying fix-inducing changes: An empirical study of apache projects},''
  {\em IEEE Transactions on Software Engineering}, vol.~48, no.~9,
  pp.~3297--3309, 2021.

\bibitem{harutyunyan2020managing}
N.~Harutyunyan, ``{Managing your open source supply chain-why and how?},'' {\em
  IEEE Computer}, vol.~53, no.~6, pp.~77--81, 2020.

\bibitem{nikitin2017chainiac}
K.~Nikitin, E.~Kokoris-Kogias, P.~Jovanovic, N.~Gailly, L.~Gasser, I.~Khoffi,
  J.~Cappos, and B.~Ford, ``Chainiac: Proactive software-update transparency
  via collectively signed skipchains and verified builds.,'' in {\em
  Proceedings of USENIX Security Symposium}, pp.~1271--1287, 2017.

\bibitem{log4j}
L.~Tal, ``{The Log4j vulnerability and its impact on software supply chain
  security}.'' Snyk Blog, 2021.

\bibitem{sbom_formats}
``{Survey of Existing SBOM Formats and Standards}.'' United States Department
  of Commerce -- National Telecommunications and Information Administration,
  2021.

\bibitem{SotoValeroMB22}
C.~Soto{-}Valero, M.~Monperrus, and B.~Baudry, ``{The Multibillion Dollar
  Software Supply Chain of Ethereum},'' {\em IEEE Computer}, vol.~55, no.~10,
  pp.~26--34, 2022.

\bibitem{MassacciP21}
F.~Massacci and I.~Pashchenko, ``Technical leverage: Dependencies are a mixed
  blessing,'' {\em {IEEE} Security \& Privacy}, vol.~19, no.~3, pp.~58--62,
  2021.

\bibitem{NTIA21}
``{The Minimum Elements For a Software Bill of Materials}.'' United States
  Department of Commerce -- National Telecommunications and Information
  Administration, 2021.

\bibitem{depclean}
C.~Soto-Valero, N.~Harrand, M.~Monperrus, and B.~Baudry, ``{A comprehensive
  study of bloated dependencies in the Maven ecosystem},'' {\em Empirical
  Software Engineering}, no.~3, pp.~1--44, 2021.

\bibitem{xia2023}
B.~Xia, T.~Bi, Z.~Xing, Q.~Lu, and L.~Zhu, ``{An Empirical Study on Software
  Bill of Materials: Where We Stand and the Road Ahead},'' in {\em Proceedings
  of ICSE}, 2023.

\bibitem{rafi2021readiness}
S.~Rafi, W.~Yu, M.~A. Akbar, S.~Mahmood, A.~Alsanad, and A.~Gumaei, ``Readiness
  model for devops implementation in software organizations,'' {\em Journal of
  Software: Evolution and Process}, vol.~33, no.~4, 2021.

\bibitem{zahan2022}
N.~Zahan, T.~Zimmermann, P.~Godefroid, B.~Murphy, C.~Maddila, and L.~Williams,
  ``{What Are Weak Links in the Npm Supply Chain?},'' in {\em Proceedings of
  ICSE-SEIP}, p.~331–340, 2022.

\end{thebibliography}
\end{document}